\documentclass[aps,prb,twocolumn,showpacs,superscriptaddress,floatfix,nofootinbib,nobibnotes]{revtex4-1}
\usepackage{amsfonts}
\usepackage{amsmath}
\usepackage{amssymb}
\usepackage{graphicx}
\usepackage{float}
\usepackage{mathtools}

\DeclareFontEncoding{LGR}{}{}
\DeclareTextSymbol{\~}{LGR}{126}
\newcommand{\lyxmathsym}[1]{\ifmmode\begingroup\def\b@ld{bold}
 \text{\ifx\math@version\b@ld\bfseries\fi#1}\endgroup\else#1\fi}

\begin{document}

\title{Effect of annealing on Raman scattering spectra of monolayer graphene samples gradually disordered by ion irradiation}

\date{\today }

\author{E. Zion}

\affiliation{Institute of Nanotechnology and Advanced Materials, Bar-Ilan University,
Ramat Gan 52900, Israel }

\author{A. Butenko}

\affiliation{Institute of Nanotechnology and Advanced Materials, Bar-Ilan University,
Ramat Gan 52900, Israel }

\author{Yu. Kaganovskii}

\affiliation{Jack and Pearl Resnick Institute, Department of Physics, Bar-Ilan
University, Ramat Gan 52900, Israel }

\author{V. Richter}

\affiliation{Institute of Nanotechnology and Advanced Materials, Bar-Ilan University,
Ramat Gan 52900, Israel }

\author{L. Wolfson}

\affiliation{Jack and Pearl Resnick Institute, Department of Physics, Bar-Ilan
University, Ramat Gan 52900, Israel  }

\author{A. Sharoni}

\affiliation{Institute of Nanotechnology and Advanced Materials, Bar-Ilan University,
Ramat Gan 52900, Israel }

\author{E. Kogan}

\affiliation{Jack and Pearl Resnick Institute, Department of Physics, Bar-Ilan
University, Ramat Gan 52900, Israel }

\author{M. Kaveh}

\affiliation{Jack and Pearl Resnick Institute, Department of Physics, Bar-Ilan
University, Ramat Gan 52900, Israel }

\author{I. Shlimak}

\email[Corresponding author: ]{ishlimak@gmail.com}
\affiliation{Jack and Pearl Resnick Institute, Department of Physics, Bar-Ilan
University, Ramat Gan 52900, Israel }

\begin{abstract}
The Raman scattering spectra (RS) of two series of monolayer graphene samples irradiated with various doses of C$^{+}$ and Xe$^{+}$ ions were measured after annealing in high vacuum, and in forming gas (95\%Ar+5\%H$_{2}$). It was found that these methods of annealing have dramatically different influence on the RS lines. Annealing in vacuum below 500$^{\circ}$C leads to significant decrease of both D-line, associated with defects, and 2D-line, associated with the intact lattice structure, which can be explained by annealing-induced enhanced doping. Further annealing in vacuum up to 1000$^{\circ}$C leads to significant increase of 2D-line together with continuous decrease of D-line, which gives evidence of partial removal of defects and recovery of the damaged lattice. Annealing in forming gas is less effective in this sense.
The blue shift of all lines is observed after annealing. It is shown that below 500$^{\circ}$C, the unintentional doping is the main mechanism of shift, while at higher annealing temperatures, the lattice strain dominates due to mismatch of the thermal expansion coefficient of graphene and the SiO$_{2}$ substrate. Inhomogeneous distribution of stress and doping across the samples leads to the correlated variation of the amplitude and the peak position of RS lines.

\end{abstract}


\maketitle

\section{introduction}

The controlled modification of graphene properties is essential for its proposed electronic applications \cite{1,2,3,4,5}. Ion irradiation is widely used for this aim (see, for example, \cite{6,7,8,9}) due to the ability to control the energy of ions and irradiation dose with high accuracy. Irradiation of pristine graphene results in increase of disorder due to introduced structural defects which influences its electrical and optical properties.

Ion irradiation as the method to introduce disorder is interesting also due to the possible reversibility caused by annealing of radiation damage. Many works were devoted to the annealing of mono- and multi-layered graphene films. However, in most of previous papers, annealing was used for pristine, non-irradiated graphene as a procedure for overcoming unintentional doping and removal of polymer residues, which remain after wet graphene transfer to the substrate or after photolithography used in the device processing \cite{10,11,12,13,14}. In a few papers, the procedure of annealing was employed to samples preliminary irradiated with ions \cite{15,16}, plasma \cite{17} and UV light \cite{18}.

Usually, measurements of the Raman scattering spectra (RS) are considered as an effective tool for probing the structure of disordered graphene films and density of introduced defects \cite{19,20,21}. Typical RS spectra for disordered graphene consist of three main lines. The G-line at 1600 cm$^{-1}$ is common for different carbon-based materials, including carbon nanotubes, mono- and multilayered graphene and graphite. The 2D-line at 2700 cm$^{-1}$ is related to an inter-valley two phonon mode, fully corresponds to momentum conservation and is emitted in the intact crystalline structure removed from any structural defects. The “defect-connected” D-line at 1350 cm$^{-1}$ is related to the inter-valley single phonon scattering process which is forbidden in the perfect graphene lattice due to momentum conservation, but is possible in the vicinity of a lattice defect (edge, vacancies, etc.) Therefore, the intensity of D-line is used (in the form of dimensionless ratio of amplitudes of D- and G-lines, $\alpha=I_{D}/I_{G}$) as a measure of disorder in graphene layers. Correspondingly, the normalized intensity of 2D-line $\beta=I_{2D}/I_{G}$ can be considered as a measure of non-destroyed part of the lattice.

In this work, we report the results of measurements of RS in monolayer graphene samples irradiated with different dose $\Phi$ of heavy (Xe) and light (C) ions and annealed at different temperatures in vacuum and in forming gas (95\%Ar+5\%H$_{2}$).

\section{samples}

Details of sample preparation, ion irradiation and measurements of RS in our samples before annealing were reported in our previous papers \cite{22, 23}. Two initial large scale monolayer graphene specimens (5x5 mm) were supplied by Graphenea Inc.. Monolayer graphene was produced by CVD on copper catalyst and transferred to a 300 nm SiO$_{2}$/Si substrate using wet transfer process. Graphene specimens of such a large size were not a monocrystalline, they look like polycrystalline films with the average size of microcrystals about 10 microns \cite{22}.

On the surface of the first specimen, six groups of micro-samples (0.2x0.2 mm) were prepared by means of electron-beam lithography. Each group of samples was irradiated by different dose $\Phi$ of carbon ions C$^{+}$ with energy of 35 keV.  On the surface of the second specimen, micro-samples were not fabricated. Six areas 2x1 mm$^{2}$ each, of the whole specimen were just irradiated by different dose $\Phi$ of Xe$^{+}$ ions with the same energy of 35 keV. As a result, two series of samples irradiated by heavy (Xe$^{+}$) and light (C$^{+}$) ions were obtained. In the RS measurements, excitation was realized by a laser beam with excitation wavelength $\lambda$ = 532 nm and power less than 2 mW to avoid heating and film destruction.

It was shown in \cite{23}, that dependences $\alpha(\Phi)$ and $\beta(\Phi)$ for both series of samples are merged if plotted not as a function of $\Phi$, but as a function of the density of defects $N_{D}$ = $k\Phi$, introduced by irradiation, where the coefficient $k$ depends on the energy and mass of the incident ion and reflects the average fraction of carbon vacancies in the graphene lattice per ion impact. It was found that for C$^{+}$-series, $k \approx 0.08$, while for Xe$^{+}$-series, $k \approx 0.8$ \cite{24}. Dependences of $\alpha(N_{D}$) and $\beta(N_{D}$) for both series of samples before annealing are shown in Fig. 1. Alignment of both dependences plotted on this scale, allows us to attribute all changes in the RS spectra observed after annealing, to the different annealing conditions.

In non-irradiated samples (for these samples we assume that $N_{D} \approx 10^{11}$ cm$^{-2}$), $\alpha$ is very small and $\beta$ is maximal. With increase of $N_{D}$, $\alpha$ increases while $\beta$ decreases. However, with further increase of $N_{D}$, $\alpha$ reaches a maximum and then decreases. This non-monotonic behavior of $\alpha$ is explained by theoretical model \cite{25} based on the assumption that a single ion impact leads to formation of completely destroyed "defective” area, $S$-area, in the immediate vicinity of the defect which is surrounded by a more extended “activated” area ($A$-area), where the graphene lattice is preserved, but the proximity to the defect causes a breakdown of the selection rules and gives rise to the emission of D-peak, attributed to a single-phonon scattering. Increasing of $A$-areas obviously results in an increase of $\alpha$ and decrease of $\beta$. However, increase of $N_{D}$ is accompanied by decrease of the mean distance between defects $L_{D} = (N_{D})^{-1/2}$, and when $L_{D}$ becomes shorter than the size of $A$-area, they begin to overlap with each other and with $S$-areas. As a result, the value of $\alpha$ reaches a maximum and then decreases. Additionally, Fig. 1 shows that $\alpha$ and $\beta \to 0$ at $N_{D} = 0.5\times10^{14}$ cm$^{-2}$. Disappearance of both Raman scattering lines could be explained by the fact that at this $N_{D}$, the mean distance between defects $L_{D} = (N_{D})^{-1/2} \approx 1.5$ nm becomes smaller than the Raman relaxation length, 2 nm \cite{26}.

Annealing of samples from Xe-series was performed in high vacuum ($2-4\times10^{-6}$ Torr), while samples from C-series were annealed in the mixed forming gas: 95\%Ar+5\%H$_{2}$ (800 sccm). Before turning on the gas flow, the tube was pumped and purged to a pressure about 100 Torr. Samples were heated at a rate of 15$^{\circ}$C/min to different annealing temperatures, $T_{a}$, and then annealed for 1 hour. Cooling of the samples was performed by shutting off the heater and letting samples cool naturally.

\begin{figure}[H]
	\includegraphics[scale=0.33]{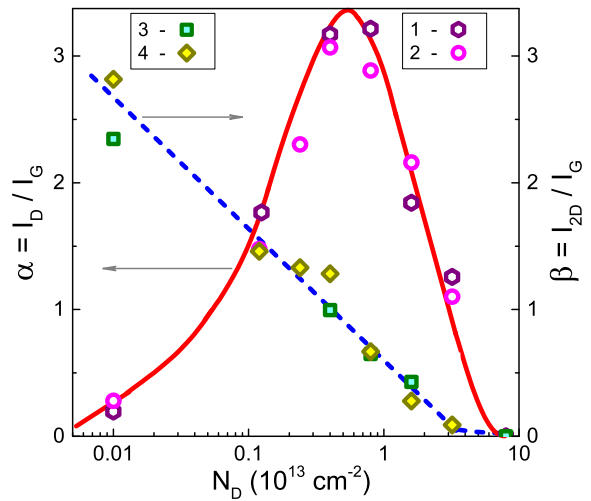}
	\caption{(Color online) Normalized amplitude of D-line $\alpha = I_{D}/I_{G}$ (1, 2) and 2D-line $\beta = I_{2D}/I_{G}$ (3, 4) for samples irradiated with different dose $\Phi$ of C$^{+}$ ions (1, 3) and Xe$^{+}$ ions (2, 4) as a function of the density of introduced defects $N_{D} = k\Phi$.  For non-irradiated samples, $N_{D}$ is presumed 10$^{11}$ cm$^{-2}$.}
	\label{Layout1}
\end{figure}

\section{results and discussion}

Figures 2 and 3 show RS measurements of samples annealed in vacuum and in forming gas, accordingly, and at different $T_{a}$. All spectra are normalized to the intensity of the G-line which is taken as 1. One can see that annealing leads to changes in amplitudes of D and 2D-lines, as well as to shifts of the frequencies of the peak positions.

\begin{figure}[H]
	\includegraphics[scale=0.35]{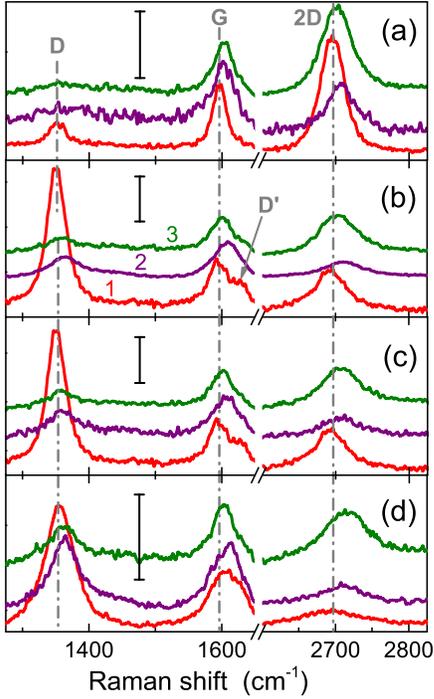}
	\caption{(Color online) Raman spectra of samples annealed in vacuum. From bottom to top: red(1) - before annealing, purple(2) and green(3) - after annealing at 550$^{\circ}$C and 1000$^{\circ}$C. $N_{D}$ (in units of 10$^{13}$ cm$^{-2}$): (a) - 0.01 (non-irradiated), (b) - 0.4, (c) - 0.8, (d) - 1.6. All spectra are shifted for clarity and normalized to the intensity of G-line, $I_{G} = 1$.}
	\label{Layout2}
\end{figure}

\begin{figure}[H]
	\includegraphics[scale=0.33]{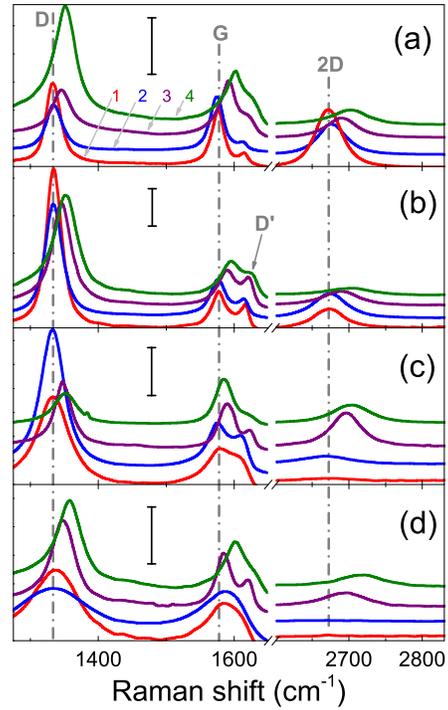}
	\caption{(Color online) Raman spectra of samples annealed in forming gas.  From bottom to top: 1 - before annealing, 2, 3, 4 - after annealing at 200$^{\circ}$C, 600$^{\circ}$C and 1000$^{\circ}$C correspondingly. $N_{D}$ (in units of 10$^{13}$ cm$^{-2}$): (a) - 0.01 (non-irradiated), (b) - 0.4, (c) - 1.6, (d) - 3.2. All spectra are shifted for clarity and normalized to the intensity of G-line, $I_{G} = 1$.}
	\label{Layout3}
\end{figure}

Figs. 4a,b show the values of $\alpha$ and $\beta$ for samples annealed in vacuum at different $T_{a}$. One can see that $\alpha$ decreases significantly with increase of $T_{a}$, with most of the change occurings at 550$^{\circ}$C. Decreasing $\alpha$ can be interpreted as a removal of irradiation induced defects. One might expect that decrease of $\alpha$ will lead to corresponding increase of $\beta$. However, Fig. 4b shows that after annealing at $T_{a} = 550^{\circ}$C, $\beta$ decreases for pristine and slightly irradiated samples, and increases only with further increase of $T_{a}$. In Ref. \cite{27}, simultaneous decrease of D- and 2D-lines was observed in defected graphene with increase of doping. So, we may suggest that in our experiment, doping is induced by vacuum annealing at low $T_{a}$ followed by exposure of annealed samples to ambient air.

Further increase of $T_{a}$ up to 1000$^{\circ}$C leads to increase of $\beta$ up to the values almost equal to those before annealing for lightly irradiated samples and up to relatively large values for strongly disordered samples where 2D-line was initially suppressed by ion irradiation (see insert in Fig. 4b). This can be attributed to the effective defect removal and partial reconstruction of the lattice structure. However, after final annealing at $T_{a}$ = 1000$^{\circ}$C, D-line still remains which means that defects are not fully removed. We note, that in samples more disordered before annealing, the values of $\alpha$ which remain after full annealing, are larger (see insert in Fig. 4a).

\begin{figure}[H]
	\includegraphics[scale=0.33]{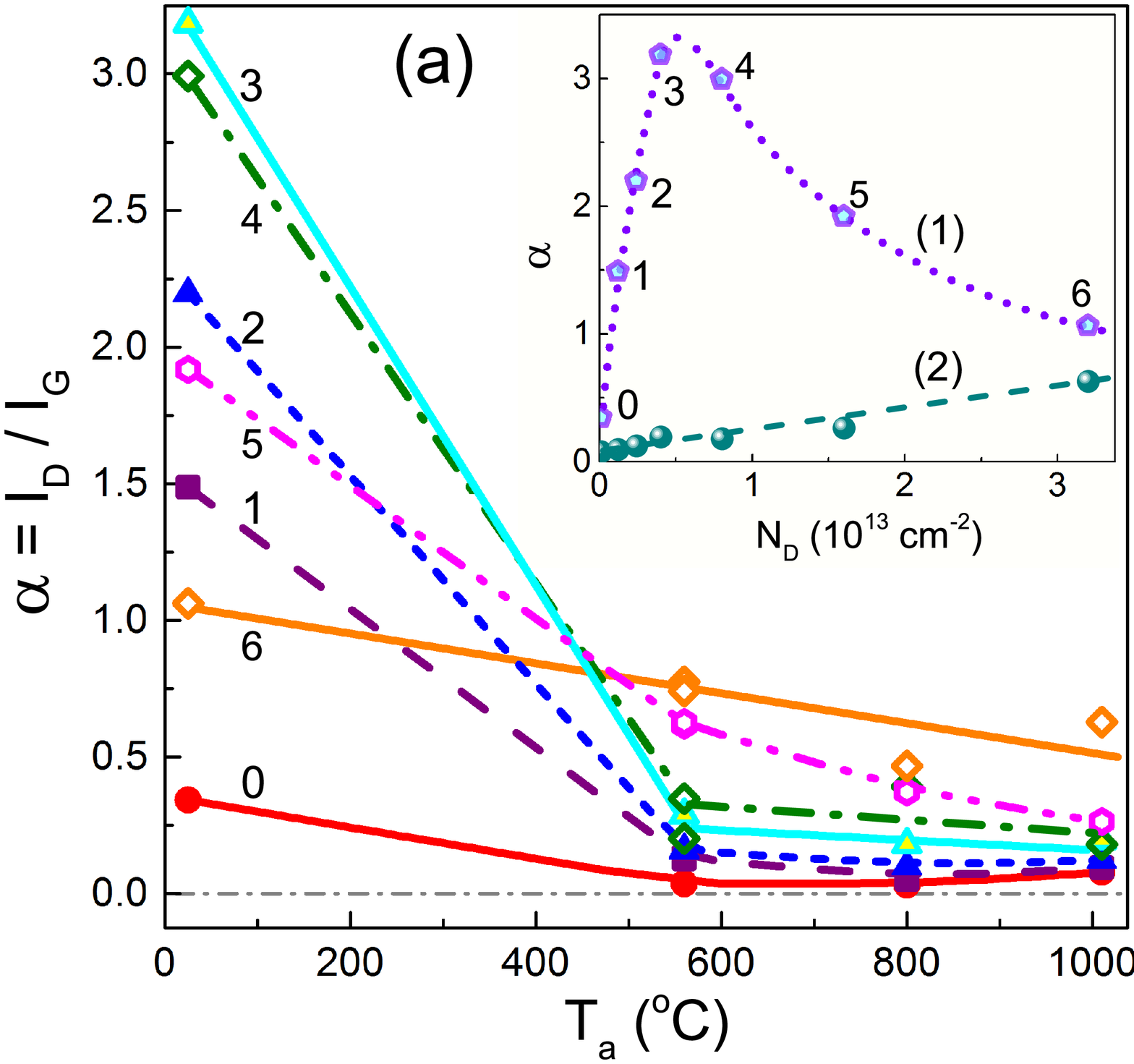}
	\includegraphics[scale=0.33]{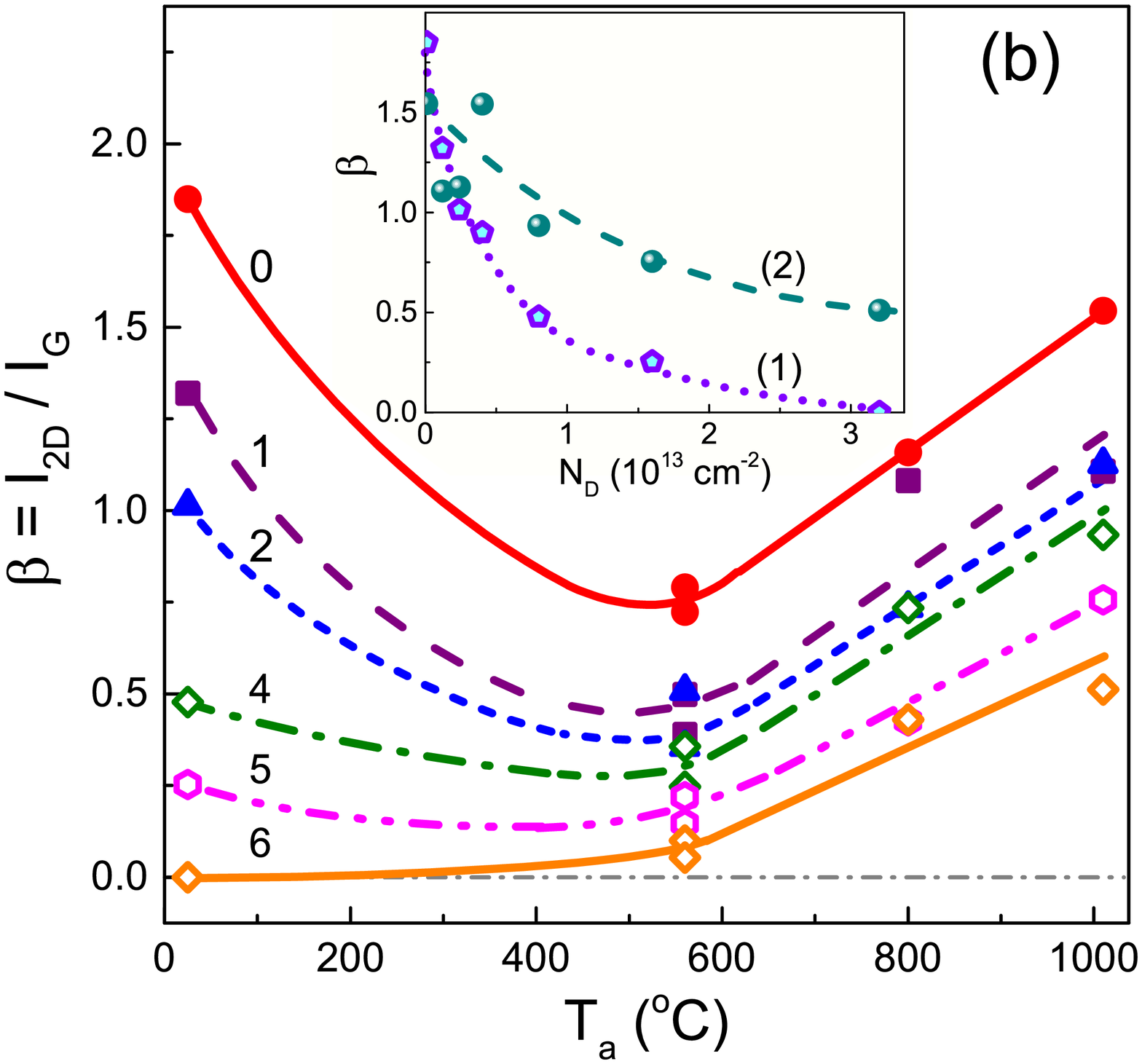}
	\caption{(Color online) $\alpha=I_{D}/I_{G}$ (a) and $\beta=I_{2D}/I_{G}$ (b) after annealing in vacuum at different temperatures $T_{a}$. The numbers near each curve correspond to the sample number shown in insert in (a). In inserts, curves (1) show $\alpha$ and $\beta$ before annealing, curves (2) – after annealing at 1000$^{\circ}$C. The lines are the guide to the eyes.}
	\label{Layout4}
\end{figure}

Figure 5a,b shows similar dependences of $\alpha$ and $\beta$ as a function of $T_{a}$ for samples annealed in forming gas. This annealing is characterized by different features: all data oscillate strongly, so that it is possible to note only the general trends: the ratio $\alpha$ decreases somewhat little for slightly disordered samples, and remains more or less the same for strongly irradiated samples, so after annealing at $T_{a}$ = 1000$^{\circ}$C all samples have a distribution $\alpha$($N_{D}$) with a maximum, which is similar than before annealing (see insert in Fig. 5a and compare with insert in Fig. 4a). The ratio $\beta$ continuously decreases a little with increase of $T_{a}$ for slightly disordered samples and increases for strongly disordered samples in which the 2D-line was initially suppressed. As a result, after final annealing at 1000$^{\circ}$C, the values of $\beta$ are small and lower than those values in the case of vacuum annealing (compare insert in Fig. 5b and insert in Fig. 4b). This gives rise to the conclusion that annealing in forming gas is less effective in reconstruction of the damaged graphene lattice.

\begin{figure}[H]
	\includegraphics[scale=0.33]{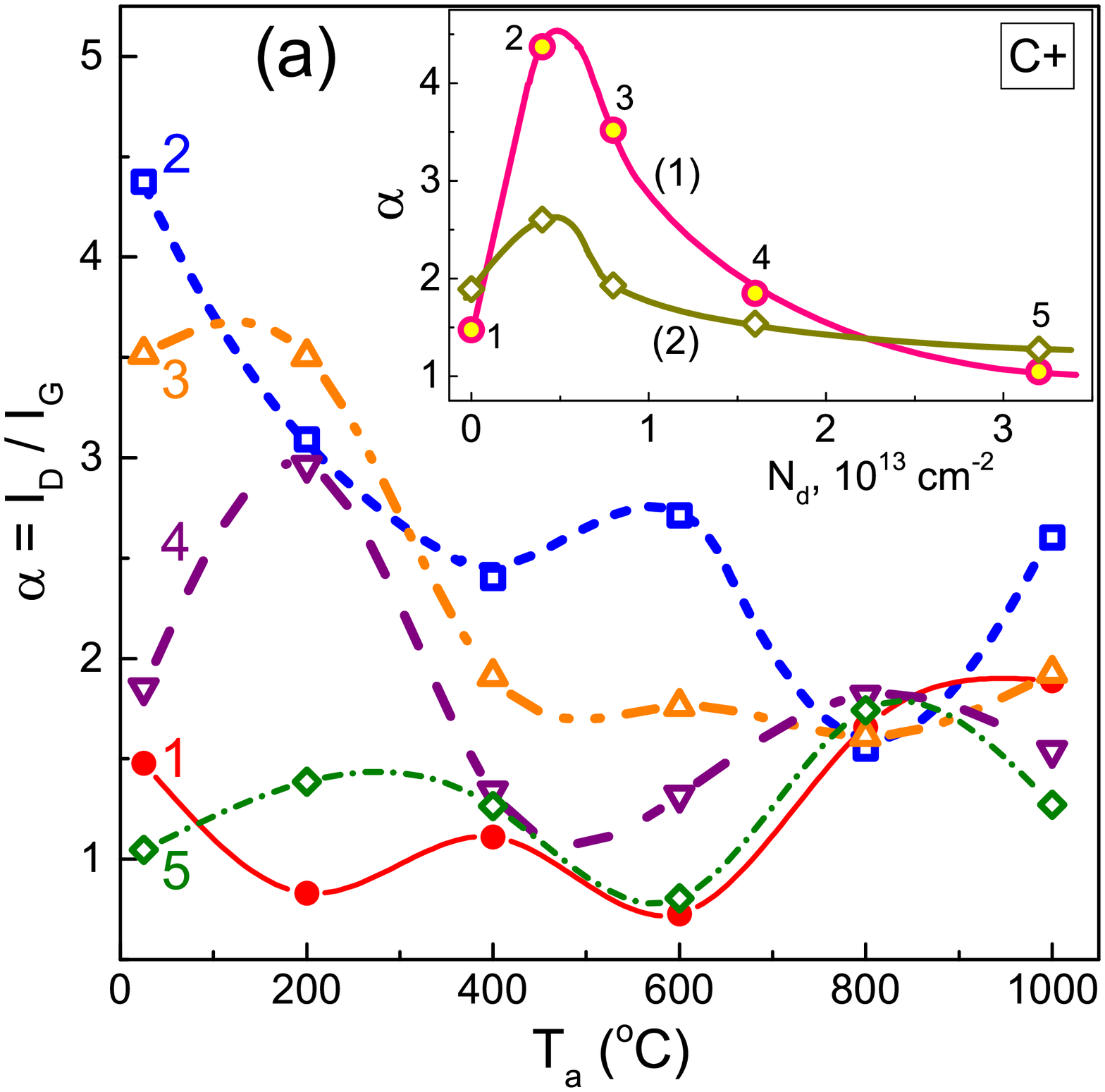}
	\includegraphics[scale=0.33]{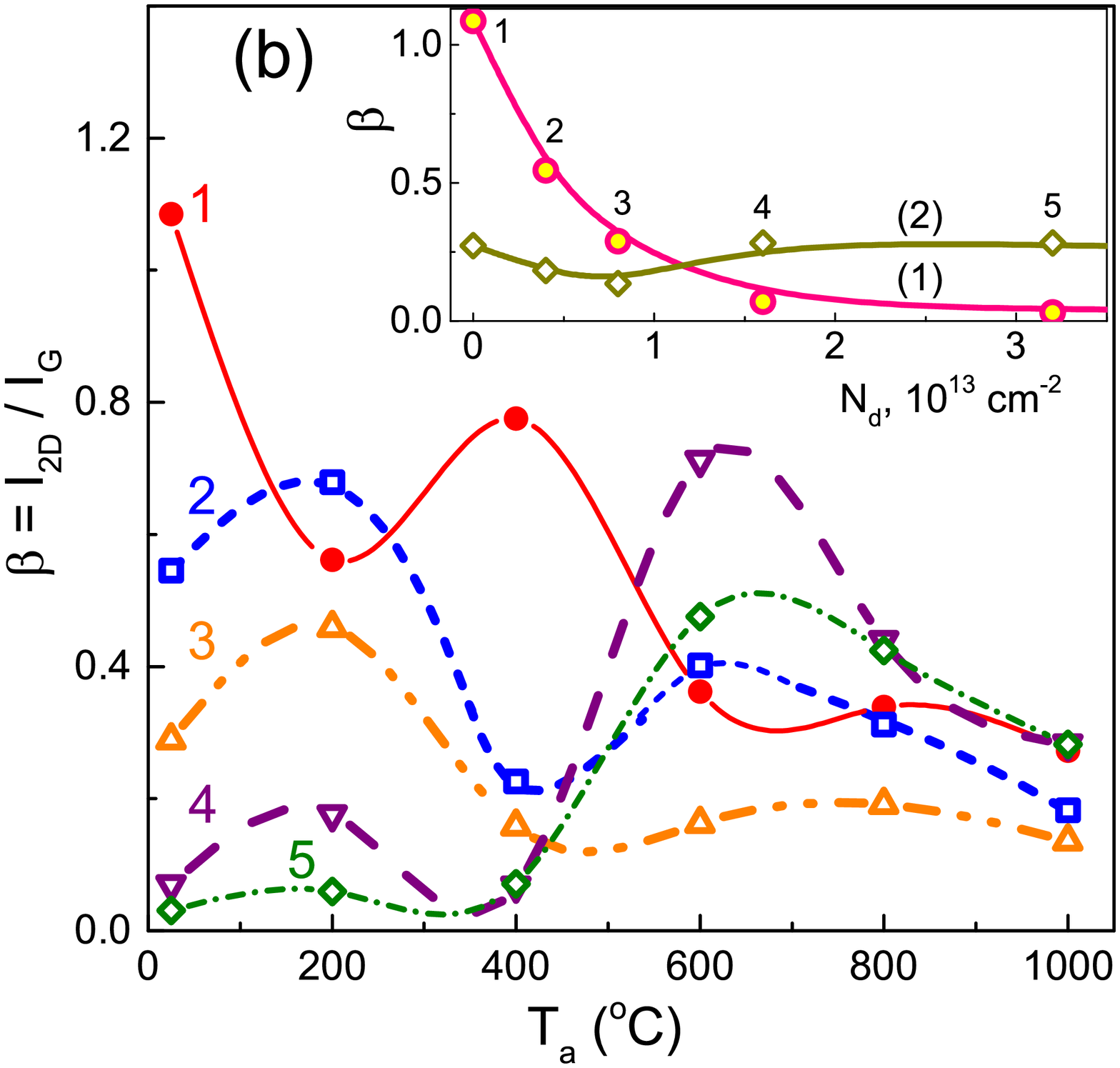}
	\caption{(Color online) $\alpha=I_{D}/I_{G}$ (a) and $\beta=I_{2D}/I_{G}$ (b) after annealing in forming gas at different temperatures $T_{a}$. The numbers near each curve correspond to the sample number shown in insert. $N_{D}$ (in units of 10$^{13}$ cm$^{-2}$):  1 - 0.01 (non-irradiated), 2 - 0.4, 3 - 0.8, 4 - 1.6, 5 - 3.2.  In inserts, curves (1) show $\alpha(N_{D})$ and $\beta(N_{D})$ before annealing, curves (2) - after annealing at 1000$^{\circ}$C. The lines are the spline interpolation.}
	\label{Layout5}
\end{figure}

Vacuum annealing results also in appearance of a broad band in RS, centered, approximately, near the position of D-line at 1300 cm$^{-1}$ (Fig. 6).  In Ref. \cite{13}, a similar band was observed in graphene annealed at 400$^{\circ}$C in oxygen-free atmosphere and was attributed to amorphous carbon (aC) on the surface of graphene \cite{28}.  This can appear because of carbonization of organic traces of PMMA polymer used in the wet transfer of CVD grown graphene film onto SiO$_{2}$/Si substrate \cite{12}. Due to existence of the aC-band, the amplitude of D-line was always obtained after subtraction of aC-band, using Lorentzian decomposition as shown in Fig. 6. In samples annealed in forming gas, aC-band was not observed, which shows that the polymer traces are efficiently removed by this annealing.

\begin{figure}[H]
	\includegraphics[scale=0.3]{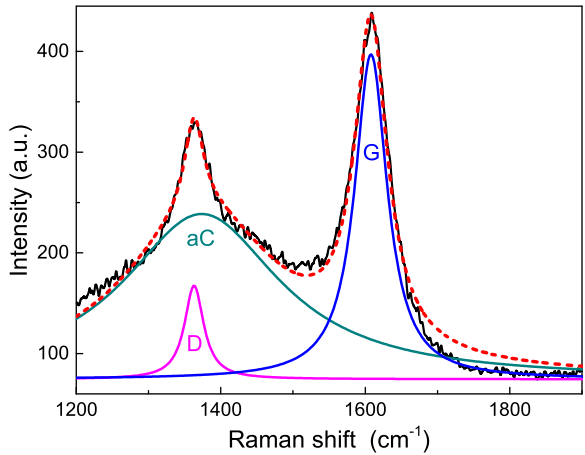}
	\caption{(Color online) Lorentzian decomposition of the Raman spectra for sample with $N_{D} = 0.4\times10^{13}$ cm$^{-2}$, annealed in vacuum at $T_{a} = 550^{\circ}$C. The black solid curve is experiment, the red dashed curve is the calculated sum of three Lorentzian curves.}
	\label{Layout6}
\end{figure}

Annealing in forming gas and in vacuum results in the blue shift of the position of all three lines, D, G and 2D. There are two reasons for the blue shift of the Raman peaks after annealing: (i) hole doping caused by enhanced ability to adsorb oxygen and water molecules after vacuum annealing and sample exposure to ambient air \cite{27,29,30} or (ii) compressive stress caused by the difference in the thermal expansion of graphene and substrate, slipping of the graphene film over substrate during heating and pinning of annealed at high $T_{a}$ film during cooling back to room temperature \cite{16,31,32,33,34,35}. It was shown in Ref. \cite{33} that these two reasons of the blue shift can be distinguished when one plots the frequency (position) of the 2D-peak $P_{2D}$ versus position of the G-peak $P_{G}$. When doping effect dominates, the slope of this dependence has to be equal to 0.7, while existence of strain leads to an increase of this slope up to 2.2. Fig. 7 shows this dependence measured at different $T_{a}$ for both series of samples. The slope is approximately 1.0 which shows that the compressive strain plays an important role in the blue shift. There is another argument to justify the contribution of lattice strain to the blue shift. The lattice deformation induced by strain leads to a change in the phonon energy, and therefore shift of 2D-line connected with emission of two equal phonons has to be double than the shift of D-line. In Fig. 8, shifts $\Delta P$ of position of RS lines after annealing at different $T_{a}$ are shown. One can see that below $T_{a}$ = 500$^{\circ}$C, shifts of all lines are equal which indicates that it is caused mainly by doping, while with further increase of $T_{a}$, shift of 2D-line is larger and the final shift after annealing at 1000$^{\circ}$C for 2D-line is, indeed, approximately two times larger than that for D-line. This shows that at high annealing temperatures, the main reason of shift is a compressive stress of the graphene lattice.

One can see from Fig. 8, that the frequency (position) of both peaks oscillate with increase of $T_{a}$. Figure 5 shows that the intensities of D- and 2D-lines also oscillate when plotted as a function of $T_{a}$. Taking into account that both changes are caused by annealing-induced lattice deformation and doping, one can expect a possible correlation between these quantities. Fig. 9 shows that, indeed, there is some correlation between variation of the intensities of RS lines and their peak positions.  In Fig. 9, the line amplitudes are normalized to the corresponding values at room temperature (RT). Such correlated variation can originate from heterogeneous distribution of the doping and the lattice strain. This can be caused by grain boundaries in the initial polycrystalline large-scale graphene specimens.

\begin{figure}[H]
	\includegraphics[scale=0.33]{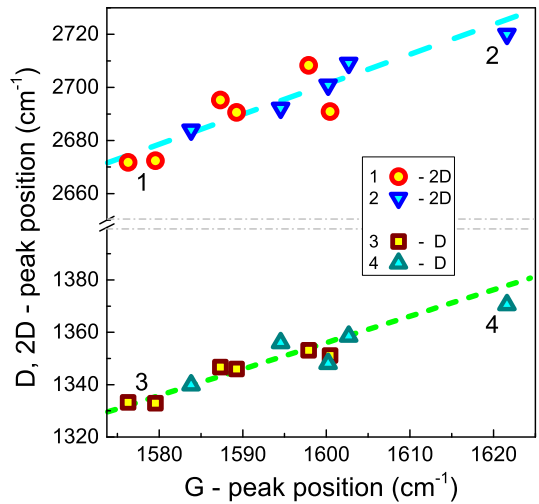}
	\caption{(Color online) Positions of 2D-peak (1, 2) and D-peak (3, 4) plotted versus position of G-peak for samples, annealed at different $T_{a}$ in forming gas (1, 3) and in vacuum (2, 4).}
	\label{Layout7}
\end{figure}

\begin{figure}[H]
	\includegraphics[scale=0.3]{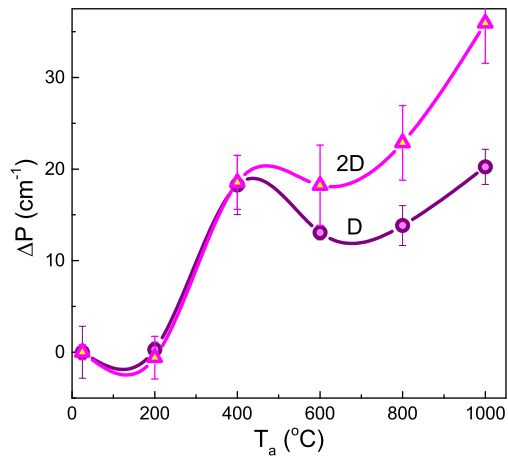}	
	\caption{(Color online) Shift of the average position of 2D-peak and D-peak for samples annealed in forming gas at different $T_{a}$. Error bars show spreading of the peak positions for samples with different $N_{D}$.}
	\label{Layout8}
\end{figure}

\begin{figure}[H]
	\begin{center}
		\includegraphics[scale=0.33]{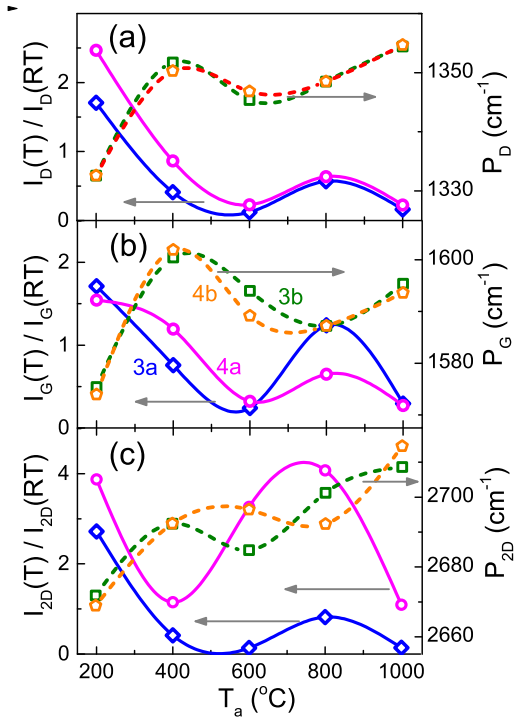}
	\end{center}
	\caption{(Color online) Normalized amplitudes (3a, 4a) and peak position (3b, 4b) of D-line (a), G-line (b) and 2D-line (c) for samples 3 and 4 annealed in forming gas at different $T_{a}$.  The amplitudes are normalized to the corresponding values at room temperature (RT) before annealing. The lines are a spline interpolation.}
	\label{Layout9}
\end{figure}

\section{conclusions}

 In conclusion, the following principal features for annealing of irradiated graphene in vacuum and in forming gas are revealed:\\
1)	Vacuum annealing below 500$^{\circ}$C, leads to simultaneous decrease of both D-line and 2D-line which can be explained by unintentional doping caused by exposure of annealed samples to ambient air. Further increase of $T_{a}$ up to 1000$^{\circ}$C leads to partial removal of defects and reconstruction of the damaged lattice. Annealing in forming gas is less effective in reconstruction of the damaged lattice up to annealing at 1000$^{\circ}$C.\\
2)	Annealing in vacuum leads also to appearance of a broad band near the position of D-line which is attributed to formation of amorphous carbon on the surface of graphene caused by carbonization of traces of polymer used in wet transfer of CVD grown graphene film on SiO$_{2}$/Si substrate. Annealing in forming gas does not lead to appearance of such a broad band, which indicates that polymer residues are removed by annealing in forming gas but only agglomerate in vacuum.\\
3)	Annealing in vacuum and in forming gas is accompanied by a blue shift of all RS lines, which is due to unintentional doping and compressive stress caused by different thermal expansion of monolayer graphene and the substrate. Doping is the main mechanism of shift at annealing below 500$^{\circ}$C, while stress dominates at high $T_{a}$ up to 1000$^{\circ}$C.\\
4)	Fluctuations of the intensity and peak position of RS lines are correlated and indicate inhomogeneous distribution of doping and strain across the samples, which may be caused by location of samples on the polycrystalline large-scale graphene specimen.

\section{acknowledgements}

E.Z ans A.S. were supported by the ISF (grant No. 569/16)

\end{document}